# The Constraints Determining Forms of Entropy and Inducing Equilibrium Distributions

Xiangjun Feng

World Chinese Forum on Science of General Systems (WCFSGS)*

*Wexford, Pennsylvania, USA*

drfeng2003@yahoo.com

*Abstract*— **Six theorems about the constraints put on general systems are established. These theorems demonstrate the decisive role of the constraints in the evolution towards equilibrium distributions, and describe the relationship between the unified constraints and the forms of entropy as the target functions for the extremization with Maximal Entropy Principle (MEP). One of the theorems justifies the fundamental postulate of statistical mechanics or the equal *a priori* probability postulate under a relative weak conditions compared with Jaynes' approach. It is shown that the constraints determining the linear functions of Boltzmann-Gibbs Entropy as the target functions are a constrained and averaged Weber-type general information which is a logarithmic function of the equilibrium distributions and can be used to directly derive various equilibrium distributions. It is also demonstrated that the constraints forcing the linear functions of Tsallis Entropy to be the target functions are a constrained and averaged Stevens-type general information which is a power-law about the equilibrium distributions and may also be applied to directly derive power-law distributions. The Tsallis entropy with q=2, called average "roundness", is shown to have a clear physical meaning. The ideas and the axiom of the constraint-based statistics are introduced. The constraint-based statistics is a further development to Jaynes' approach, and pays a great emphasis on the constraints instead of on the entropy.**

## I. INTRODUCTION

THERE has been a tremendous interest in a generalized entropy or Tsallis entropy [1]. Tsallis' approach, however, has also stirred a heavy debate [2]. The reason why Tsallis entropy has attracted so much interest is partly because of the possibility to derive power-law distributions in the canonical ensemble within the maximum entropy principle [3]. Hanel and Thurner, however, showed that within classical statistical mechanics it is possible to naturally derive power-law distributions which are of Tsallis type. The only assumption is that microcanonical distributions have to be separable from the total system energy. They demonstrated with a separation technique that all separable distributions are parameterized by a separation constant Q which is one to one related to the q-parameter in Tsallis distributions, and is associated with the system size and the parameter depicting the interaction for some realistic system with interactions. The power-laws obtained are formally equivalent to those obtained by maximizing Tsallis entropy under q constraints [3]. Starting from microcanonical basis with the principle of equal *a priori* probability, Abe and Rajagopal found that, besides ordinary Boltzmann-Gibbs theory with the exponential distribution, a theory describing systems with power-law distributions can also be derived [4]. Abe and Rajagopal paid an emphasis on that they made no initial assumptions on the definition of statistical expectation value and the form of the entropy. Rostovtsev , on the other hand, argued that for a vast class of phenomena having a power-law statistics an arithmetic mean is not defined from the first principles, while an application of a geometric mean constraint is rather logic [5]. Rostovtsev's argument suggests that the traditional Boltzmann-Gibbs entropy is more suitable for the derivation of the power-law distributions. Hanel and Thurner pointed out with a somewhat different opinion that many statistical systems in nature can not be satisfactorily described by naive or straight forward application of Boltzmann-Gibbs statistical mechanics. They derived generalized Boltzmann factors and generalized Boltzmann-Gibbs entropy [6]. Another interesting work related with constraints was done by Niven who used the Lagrangian technique to derive constrained forms of the Tsallis entropy and Shannon entropy [7].

In this work, I adopt a somewhat different strategy and derive. I pay great emphasis on the constraints which are usually testable and are demonstrated directly by nature. The constraints, however, were taken as only an "additional conditions" not as important as the entropy or the equal a *priori* probability postulate for many people. An axiom of the constraint-based statistics is put forward. Six mathematical theorems are established to show that the unified forms of constraints when working together with the Maximal Entropy Principle (MEP) as an extremum principle not only fully determine the forms of the entropy but also can be used to directly derive various distributions including power-law distributions. Unified mathematical expressions of constraints are thus not at all "less important" than a generalized entropy like Tsallis entropy. The testable constraints are taken as a constrained Average Characteristic Information which induces various equilibrium distributions other than the uniform distribution corresponding to the natural constraint. The equal *a priori* probability postulate is just a direct result caused by the natural constraint working with the Maximal Entropy Principle(MEP) under a relatively weak conditions compared with Jaynes' approach. I would like to argue that a key criteria to judge if a new definition of entropy like Tsallis entropy is useful may come from the reality of the unified constraints which fully determine the entropy as the target function. I also think that an understanding to the unified constraints from a viewpoint of general information theory may help understand the physical and mathematical characteristics of new definitions of entropy like Tsallis entropy. The main purpose of this paper, however, is to establish a solid mathematical foundation for the constraint-based statistics.

## II.  THE AXIOM OF THE CONSTRAINT-BASED STATISTICS

As we know that the fundamental postulate of statistical mechanics is the equal *a priori* probability postulate. In the constraint-based statistics, I put forward the following axiom.

*The Axiom of Constraint-Based Statistics*

*Both the equilibrium distributions and the forms of entropy as the target functions are a result caused by the constraints working together with extremum principles like Maximal Entropy Principle (MEP).*

In the following sections, I will establish six theorems centered around the axiom to show the vital importance of the constraints. I will also justify the equal a *priori* probability postulate under a weak conditions. It will be shown that the equal *a priori* probability is just the result caused by the natural constraint working together with the MEP. The conditions satisfied by various forms of entropy are relatively weak.

## III.  THE UNIFIED CONSTRAINTS DETERMINING FULLY THE LINEAR FUNCTIONS OF BOLTZMANN-GIBBS ENTROPY AS THE TARGET FUNCTIONS

*Theorem 1[8][9][10]*

Assume $A$ is a finite set of $\{A_1, A_2..., A_n\}$ and is a source alphabet, $p$ is the probability set of $\{p_1, p_2..., p_n\}$ assigned to $A$, and the probability set $p$ has an equilibrium state of the set $f$ which can be expressed as $\{f_1, f_2,..., f_n\}$. Then, the expression of the entropy as the target function for the extremization with Maximal Entropy Principle (MEP) should be and should only be $aH + b$, with $a$ and $b$ as constants and $a > 0$, if the constraint put on $A$ is as follows:

$$-\sum_{i=1}^{n} p_i \ln(f_i) = \text{constant} \qquad (1)$$

In other words, the left hand side of Eq. (1) does not change with the probability set p. The above-mentioned $H$ is the normalized Boltzmann-Gibbs entropy or Shannon entropy as its mathematical or conceptual form, or

$$H = -\sum_{i=1}^{n} p_i \ln(p_i) \qquad (2)$$

I define $-\ln(f_i), \forall i$ as the *Characteristic Information* contained within the constraint about the equilibrium distribution $f$, and define $-\sum_{i=1}^{n} p_i \ln(f_i)$ as the *Average Characteristic Information [8]*. The constraints expressed by Eq. (1) for various equilibrium distributions of $f$ are thus a constrained Average Characteristic Information. The constrained Average Characteristic Information not only induces the equilibrium distributions of $f$, but also determines the linear functions of the Boltzmann-Gibbs entropy as the target functions.

There is a Weber–Fechner law which describes the relationship between the physical magnitudes of stimuli and the perceived intensity of the stimuli [11]. The mathematical form of Weber–Fechner law is as follows

$$P = k \ln(\frac{s}{s_0})$$

where, $s$ is the stimulus, $s_0$ is the stimulus resulting in no perception, $k$ is a constant, and $P$ is the perception. Let us assume that $\max f_i$ is the maximal value for all possible $f_i$, and $s_0 = \frac{1}{\max f_i} = \frac{1}{1} = 1$, $k = 1$, and $s = \frac{1}{f_i}$. We thus obtain $P = -\ln(f_i), \forall i$.

Obviously the perception following Weber-Fechner law is a general information contained within the stimulus. I call any general information described by the logarithmic function a Weber-type general information. Therefore, the Characteristic Information determining linear functions of Boltzmann-Gibbs Entropy as the target functions is a Weber-Type general information. A deep understanding about the perception-constraint analogy will be discussed elsewhere.

*Proof of Theorem 1*

With the assumption of $-\sum_{i=1}^{n} p_i \ln(f_i)$ = constant and the natural constraint of $\sum_{i=1}^{n} p_i = 1$, It is true that for any nonzero constant $c_1$ and any constant $c_2$,

$$-c_1 \sum_{i=1}^{n} p_i \ln(f_i) + c_2 \sum_{i=1}^{n} p_i = \text{constant} \quad (3)$$

One may apply the Lagrange method of undetermined multipliers to the undetermined entropy $S$ subject to the above constraint and the natural constraint. The Lagrangian $L$ can be expressed as

$$L = S + \lambda c_1 \sum_{i=1}^{n} p_i \ln(f_i) - (\lambda c_2 + \mu) \sum_{i=1}^{n} p_i \quad (4)$$

where $S$ is the undetermined entropy, and $\lambda$ and $\mu$ are undetermined multipliers. Extremization of Eq. (4) gives the maximal entropy $S$. Setting $\frac{\partial L}{\partial p_i} | (p_i = f_i) = 0, \forall i$ yields the equations resulting in the maximal entropy (equilibrium) distribution $f$:

$$\frac{\partial S}{\partial p_i} | (p_i = f_i) = -\lambda c_1 \ln(f_i) + \lambda c_2 + \mu, \forall i \quad (5)$$

For given values of arbitrary positive constant $a$ and constant $b$, let us choose $c_1$ and $c_2$ and make them satisfy

$$c_1 = \frac{a}{\lambda} \text{ and } c_2 = \frac{(b-a-\mu)}{\lambda}$$

Since $\frac{\partial H}{\partial p_i} = -\ln(p_i) - 1$, it becomes obvious that if

$S = aH + b = aH + b\sum_{i=1}^{n} p_i$, Eq. (5) will be satisfied. If there is another expression of $S = S^*$ which also satisfies Eq. (5), one always has $S = g(aH + b)$, because $S^* = (S^*/(aH+b))(aH+b) = g(aH+b)$.

With Eq. (5),

$$\frac{\partial g}{\partial p_i}(aH + b\sum_{i=1}^{n} p_i) | (p_i = f_i) + g(-a\ln(f_i) + b - a)$$

$$= -a\ln(f_i) + b - a, \forall i$$

Since $H$ in the above equations is related with the whole set of probability p and is not only associated with a specific probability of $p_i$ which equals to $f_i$, the only possibility to make the equations true is to let $\frac{\partial g}{\partial p_i} = 0, \forall i$, and $g = 1$.

As a summary, the entropy $S$ should be and should only be with an expression of $aH + b$ where $H$ is the normalized Boltzmann-Gibbs entropy. In order to guarantee that $S$ becomes maximal when $H$ becomes maximal through the extremization, the constant $a$ should also be positive. I have completed the proof.

For classic extensive systems, the Boltzmann-Gibbs entropy or Shannon entropy (H) is directly related with the logarithm of the probability of occurrence [15]. Therefore, the unified constraints described by Eq.(1) are with a maximal probability of occurrence.

## IV. THE UNIFIED CONSTRAINTS DETERMINING FULLY THE LINEAR FUNCTIONS OF TSALLIS ENTROPY AS THE TARGET FUNCTIONS

*Theorem 2*

Assume $A$ is a finite set of $\{A_1, A_2..., A_n\}$ and is a source alphabet, $p$ is the probability set of $\{p_1, p_2..., p_n\}$ assigned to $A$, and the probability set $p$ has an equilibrium state of the set $f$ which can be expressed as $\{f_1, f_2,..., f_n\}$. Then, the expression of the entropy as the target function for the extremization with the Maximal Entropy Principle (MEP) should be and should only be $aS_q + b$, with $a$ and $b$ as constants and $a > 0$, if the constraint put on A is as follows:

$$\sum_{i=1}^{n} p_i f_i^{(q-1)} = \text{constant} \quad (6)$$

In other words the left hand side of Eq. (6) does not change with the probability set $p$. The above-mentioned $S_q$ is the Tsallis entropy expressed in the form as a statistical average.

$$S_q = \frac{1}{(q-1)} \sum_{i=1}^{n} p_i (1 - p_i^{(q-1)}) \quad (7)$$

I define $f_i^{(q-1)}, \forall i$ as the *Characteristic Information* contained within the constraint about the equilibrium distribution $f$, and define $\sum_{i=1}^{n} p_i f_i^{(q-1)}$ as the *Average Characteristic Information*. The unified constraints expressed by Eq. (6) are thus a constrained Average Characteristic Information. The constrained Average Characteristic Information not only induces the equilibrium distributions of $f$, but also determines the linear functions of the Tsallis entropy as the target functions.

It should be mentioned that when $q = 2$, the Average Characteristic Information is just the arithmetic mean of the equilibrium probabilities.

There is another law called Stevens' power-law which also describes the relationship between the magnitude of a physical stimulus and its perceived intensity or strength [12]. The mathematical expression of Stevens' power-law is as follows.

$$\psi(I) = kI^a$$

where $I$ is the magnitude of the physical stimulus, $\psi$ is the psychophysical function capturing sensation (the subjective size of the stimulus), $a$ is an exponent that depends on the type of stimulation and $k$ is a proportionality constant that depends on the type of stimulation and the units used. Assume $I = \frac{1}{f_i}$, $a = 1-q$, $k = 1$, one obtains

$$\psi(fi) = f_i^{(q-1)}, \forall i$$

Obviously the psychophysical function capturing sensation describes a type of general information contained within the physical stimulus. I call any general information following the power-law a Stevens-type general information. Therefore, the Characteristic Information determining linear functions of Tsallis Entropy as the target functions is a Stevens-type general information. The analogy between psychophysical function and Average Characteristic Information will be studied elsewhere.

*Proof of Theorem 2*

The proof is similar to the proof of the theorem1.

With the assumption of $\sum_{i=1}^{n} p_i f_i^{(q-1)}$ = constant and the natural constraint of $\sum_{i=1}^{n} p_i = 1$, it is true that for any nonzero constant $c_1$ and any constant $c_2$,

$$c_1 \sum_{i=1}^{n} p_i f_i^{(q-1)} + c_2 \sum_{i=1}^{n} p_i = \text{constant} \quad (8)$$

One can apply the Lagrange method of undetermined multipliers to the undetermined entropy $S$ subject to the above constraint and the natural constraint. The Lagrangian $L$ can be expressed as

$$L = S - \lambda c_1 \sum_{i=1}^{n} p_i f_i^{(q-1)} - (\lambda c_2 + \mu) \sum_{i=1}^{n} p_i \quad (9)$$

where $S$ is the underdetermined entropy, and $\lambda$ and $\mu$ are undetermined multipliers. Extremization of Eq. (9) gives the maximal entropy $S$. Setting $\frac{\partial L}{\partial p_i}|(p_i = f_i) = 0, \forall i$ yields the equations resulting in the maximal entropy (equilibrium) distribution $f$:

$$\frac{\partial S}{\partial p_i}|(p_i = f_i) = \lambda c_1 f_i^{(q-1)} + (\lambda c_2 + \mu), \forall i \quad (10)$$

For given values of arbitrary positive constant $a$, constant $b$, and q, let us choose $c_1$ and $c_2$, and make them satisfy:

$$c_1 = \frac{aq}{(1-q)\lambda} \text{ and } c_2 = \frac{a}{(q-1)\lambda} + \frac{(b-\mu)}{\lambda}$$

Since $\frac{\partial S_q}{\partial p_i} = \frac{1}{(q-1)} - \frac{q}{(q-1)} p_i^{(q-1)}$

It becomes obvious that if $S = aS_q + b = aS_q + b\sum_{i=1}^{n} p_i$, Eq. (10) will be satisfied. Assume that there is another expression of $S = S*$ which also satisfies Eq. (10), One always has

$S = g(aS_q + b)$, because

$$S* = (S*/(aS_q + b))(aS_q + b) = g(aS_q + b)$$

With Eq. (10), one has

$$\frac{\partial g}{\partial p_i}(aS_q + b\sum_{i=1}^{n} p_i)|(p_i = f_i) + ...$$

$$+ g(\frac{aq}{(1-q)} f_i^{(q-1)} + \frac{a}{(q-1)} + b)$$

$$= \frac{aq}{(1-q)} f_i^{(q-1)} + \frac{a}{(q-1)} + b, \forall i$$

Since $S_q$ in the above equations is related with the whole set of probability p and is not only associated with a specific probability of $p_i$ which equals to $f_i$, the only possibility to make the equations true is to let $\frac{\partial g}{\partial p_i} = 0, \forall i$ and $g = 1$.

As a summary, the entropy $S$ should be and should only be with an expression of $aS_q + b$, where $S_q$ is the Tsallis entropy. In order to guarantee that $S$ becomes maximal when the Tsallis entropy becomes maximal through the extremization, the constant $a$ should also be positive. I have completed the proof.

An illustration will be given to the theorem 1 and theorem 2 before I prove a more general theorem. I will take the power-law distributions as an example.

## V. ILLUSTRATING THE THEOREM 1 AND THEOREM 2 WITH POWER-LAW DISTRIBUTIONS

In the very first paper on the Tsallis formalism, Tsallis derived the power-law equilibrium distributions in the canonical ensemble within the maximum entropy principle[1][13].

$$f_i = [\frac{(1-q)}{q}(\alpha + \beta H_i)]^{\frac{1}{(q-1)}}, \forall i \qquad (11)$$

where $H_i$ is the internal energy for the ith state.

For such a form of the power-law, The expression of the constraints which fully determine the linear functions of Boltzmann-Gibbs Entropy as the target functions, according to the theorem 1, is $-\sum_{i=1}^{n} p_i \ln(f_i) = $ constant

With Eq. (11), the above equation is led to

$$\sum_{i=1}^{n} p_i \ln(\alpha H_i + \beta) = \text{constant} \qquad (12)$$

On the other hand, from the conclusion of the theorem 2, the expression of the constraints which fully determine the linear functions of Tsallis Entropy as the target functions is

$$\sum_{i=1}^{n} p_i f_i^{(q-1)} = \text{constant}$$

With Eq. (11), the above equation gives

$$\sum_{i=1}^{n} p_i H_i = \text{constant} \qquad (13)$$

It should be mentioned that Tsallis made the same assumption as shown in Eq. (13) [1].

From Eq. (12) and Eq. (13), one may understand that both Boltzmann-Entropy and Tsallis-Entropy can be used to derive the power-law distributions. The key is that the corresponding constraints are quite different. If the reality in nature is a constrained geometrical mean as argued by Rostovtsev [5], one has to apply the Boltzmann-Gibbs entropy to obtain the power-law distributions. However, if a realistic constraint in nature is found to be a constrained arithmetic mean, one must use Tsallis entropy to derive the power-law distributions.

## VI. WITH REAL CONSTRAINTS IN NATURE TO DIRECTLY DERIVE THE EQUILIBRIUM DISTRIBUTIONS [8][9][10]

*Theorem 3*

Assume $A$ is a finite set of $\{A_1, A_2..., A_n\}$ and is a source alphabet, $p$ is the probability set of $\{p_1, p_2..., p_n\}$ assigned to $A$, and the probability set $p$ has an achievable state of the set $f$ which can be expressed as $\{f_1, f_2,..., f_n\}$. Then the equilibrium distribution should be and should only be $f$ if the expression of the entropy as the target function for the extremization with Maximal Entropy Principle (MEP) is $aH + b$, with $a$ and $b$ as constants and $a > 0$, and if the constraint put on $A$ is as follows:

$$-\sum_{i=1}^{n} p_i \ln(f_i) = \text{constant}.$$

The above mentioned $H$ is the normalized Boltzmann-Gibbs entropy or Shannon entropy as its mathematical or conceptual form,

$$H = -\sum_{i=1}^{n} p_i \ln(p_i)$$

*Theorem 4*

Assume $A$ is a finite set of $\{A_1, A_2..., A_n\}$ and is a source alphabet, $p$ is the probability set of $\{p_1, p_2..., p_n\}$ assigned to $A$, and the probability set $p$ has an achievable state of the set $f$ which can be expressed as $\{f_1, f_2,..., f_n\}$. Then the equilibrium distributions should be and should only be $f$ if the expression of the entropy as the target function for the extremization with Maximal Entropy Principle (MEP) is $aS_q + b$, with $a$ and $b$ as constants and $a > 0$, and if the constraint put on $A$ is as follows:

$$\sum_{i=1}^{n} p_i f_i^{(q-1)} = \text{constant}.$$

The above mentioned $S_q$ is the Tsallis entropy expressed in the form as a statistical average:

$$S_q = \frac{1}{(q-1)} \sum_{i=1}^{n} p_i (1 - p_i^{(q-1)})$$

The proof of the theorem 3 and theorem 4 is very similar to the proof of the theorem 1 and 2. Therefore I omit the proof. The original proof can be found partly in [8][9][10].

I have actually used the theorem 3 to derive common types of equilibrium distributions. Let me show some of the derivations here.

*1) Uniform Distributions*

If the source alphabet A is constrained only with the natual constraint , What is the equilibrium distribution of A? Let me assume that the constraint is a Weber-type constraint with a unified expression of that

$$-\sum_{i=1}^{n} p_i \ln(f_i) = \text{constant}$$

where $f = \{f_1, f_2, ... f_n\}$ is an achievable state of the probability set p assigned to A, and the corresponding entropy as the target function for the extremization with the MEP is Boltzmann-Gibbs entropy. Then, the equilibrium distribution must be $f$, according to the theorem 3.

Let us examine the specific set of probability $f$ which can change the above general form of constraints described by Eq. (1) into the special case of the natural constraint or $\sum_{i=1}^{n} p_i = 1$. There is no other choice but that

$\ln(f_i) = \text{constant}, \forall i$

or

$f_i = c, \forall i$

Since $\sum_{i=1}^{n} f_i = 1$, one obtains

$f_i = 1/n, \forall i$

In other words, the uniform distribution is derived.

*2) Exponential Distributions*

If the source alphabet A is of a constrained arithmetic mean, or $\sum_{i=1}^{n} p_i A_i = \text{constant}$, what is the equilibrium distribution?

Let me assume the constraint is a Weber-type constraint or

$$-\sum_{i=1}^{i=n} p_i \ln(f_i) = \text{constant}$$

where $f = \{f_1, f_2, ... f_n\}$ is an achievable state of the probability set p assigned to A, and the corresponding entropy as the target function is Boltzmann-Gibbs entropy. Then, the equilibrium distribution must be $f$, according to the theorem 3. Let us examine the specific set of probability $f$ which can change the above general form of constraints described by Eq. (1) into a constrained arithmetic mean. There is no other choice but that

$-\ln(f_i) = aA_i + b, \forall i$

or

$f_i = \exp(-aA_i - b) = c \exp(-aA_i), \forall i$

where

$c = \exp(-b)$.

Since $\sum_{i=1}^{n} f_i = 1$, one has

$$c = \frac{1}{\sum_{i=1}^{n} \exp(-aA_i)} = 1/Z$$

where $Z$ is the partition function,

$$Z = \sum_{i=1}^{n} \exp(-aA_i)$$

Assume $\sum_{i=1}^{n} p_i A_i = \overline{m}$, One finally obtains

$$\overline{m} = -\frac{\partial \ln(Z(a))}{\partial a}$$

$$f_i = \frac{\exp(-aA_i)}{Z}, \forall i.$$

*3) Power-Law Distributions Derived with a Weber-Type Constraint*

If the source alphabet A is of a constrained geometrical mean, or $\sum_{i=1}^{n} p_i \ln(A_i) = \text{constant}$, what is the equilibrium distribution?

Let me assume the constraint is a Weber-type constraint or

$$-\sum_{i=1}^{i=n} p_i \ln(f_i) = \text{constant}$$

where $f = \{f_1, f_2, ... f_n\}$ is an achievable state of the probability set p assigned to A, and the corresponding entropy as the target function is Boltzmann-Gibbs entropy. Then, the equilibrium distribution must be $f$, according to theorem 3.

Let us examine the specific set of probability $f$ which can change the above general form of the constraints described by Eq. (1) into a constrained geometrical mean. There is no other choice but that $f_i = cA_i^a, \forall i$

Since $\sum_{i=1}^{n} f_i = 1$, one has

$$c = \frac{1}{\sum_{i=1}^{n} A_i^a} = \frac{1}{Z_p}$$

where $Z_p = \sum_{i=1}^{n} A_i^a$

By defining $\bar{g} = \sum_{i=1}^{n} p_i \ln(A_i)$ =constrained geometrical mean, one finally obtains:

$$\bar{g} = \frac{\partial \ln(Z_p(a))}{\partial a}$$

$$f_i = \frac{A_i^a}{Z_p}, \forall i$$

*4) Power-Law Distributions Derived with Stevens-Type Constraints*

If the source alphabet A is of a constrained arithmetic mean, or $\sum_{i=1}^{n} p_i A_i$ =constant, what is the equilibrium distribution? Let me assume the constraint is a Stevens-type constraint or

$$\sum_{i=1}^{i=n} p_i f_i^{(q-1)} = \text{constant}$$

where $f = \{f_1, f_2,...f_n\}$ is an achievable state of the probability set p assigned to A, and the corresponding entropy as the target function is Tsallis entropy. Then, the equilibrium distribution must be $f$, according to the theorem 4.

Let us examine the specific set of probability $f$ which can change the above general form of constraints described by Eq. (6) into a constrained arithmetic mean. There is no other choice but that

$$f_i = (aA_i + b)^{\frac{1}{(q-1)}}, \forall i$$

Let us assume

$$\bar{m} = \sum_{i=1}^{n} p_i A_i = \sum_{i=1}^{n} f_i A_i$$

and

$$\hat{g} = \sum_{i=1}^{n} f_i f_i^{(q-1)} = \sum_{i=1}^{n} f_i^q$$

Since

$$b = f_i^{(q-1)} - aA_i$$

multiplying both sides of the above equation with $f_i$ and taking the sum for all $i$ yields

b = $\hat{g} - a\bar{m}$

The partition function can also be found [13].

*5) What is the criteria for judging if a generalized entropy is useful?*

It is clear by now that both Boltzmann-Gibbs entropy and Tsallis Entropy can be used to derive equilibrium distributions with different constraints. Therefore, there is a reason to believe that a key criteria for judging if Tsallis entropy is useful may come from the reality of the constraints. For the power–law distributions, a key to show the power of Tsallis entropy is to find a case when the equilibrium power-law distribution is formed with a constrained arithmetic mean.

### VII. A GENERALIZED EQUATION FOR THE CONSTRAINTS FULLY DETERMINING THE ENTROPY AS THE TARGET FUNCTION

*Theorem 5*

Assume $A$ is a finite set of $\{A_1, A_2..., A_n\}$ and is a source alphabet, $p$ is the probability set of $\{p_1, p_2..., p_n\}$ assigned to $A$, and the probability set $p$ has an equilibrium state of the set $f$ which can be expressed as $\{f_1, f_2,..., f_n\}$. Then the expression of the entropy as the target function for the extremization with MEP should be and should only be $S$, if the constraint put on A is as follows:

$$-\sum_{i=1}^{n} p_i g(f_i) = \text{constant} \quad (14)$$

where

$$g(f_i) = \frac{1}{\lambda}\left(F(f_i) + f_i \frac{\partial F}{\partial p_i}\bigg|(p_i = f_i) + \mu\right) \quad (15)$$

and

$$S = -\sum_{i=1}^{n} p_i F(p_i) \quad (16)$$

In Eq. (15), $\lambda$ and $\mu$ are undetermined multipliers, and in Eq. (16), $F(p_i)$ is only determined by the specific $p_i$.

*Proof of Theorem 5*

With the assumption of $-\sum_{i=1}^{n} p_i g(f_i)$ = constant, one can apply the Lagrange method of underdetermined multipliers to an underdetermined entropy S subject to the above constraints and the natural constraint:

$$\sum_{i=1}^{n} p_i = 1 \quad (17)$$

The Lagrangian thus can be expressed as follows.

$$L = S + \lambda \sum_{i=1}^{n} p_i g(f_i) - \mu \sum_{i=1}^{n} p_i \quad (18)$$

where $S$ is the undetermined entropy, and $\lambda$ and $\mu$ are undetermined multiplies. Extremization of Eq. (18) gives the maximal entropy S.

Setting $\frac{\partial L}{\partial p_i}|(p_i = f_i) = 0, \forall i$ and recalling Eq. (16) yields the equations resulting in the maximal entropy (equilibrium) distribution f:

$$-F(f_i) - f_i \frac{\partial F}{p_i}|(p_i = f_i) = -\lambda g(f_i) + \mu \quad (19)$$

From Eq. (19), one obtains

$$g(f_i) = \frac{1}{\lambda}(F(f_i) + f_i \frac{\partial F}{\partial p_i}|(p_i = f_i) + \mu)$$

In other words, I have proven that as long as the constraint satisfies Eq. (14) and Eq. (15), The entropy described by Eq. (16) will be the target function. If there is another function $S^*(p) = g^*S(p)$, which can be used to replace the $S$ in Eq. (18), one has

$$\frac{\partial g^*}{\partial p_i}S|(p_i = f_i) + g^*(-F(f_i) - f_i \frac{\partial F}{\partial p_i}|(p_i = f_i))$$
$$= -\lambda g(fi) + \mu$$
$$= -F(f_i) - f_i \frac{\partial F}{\partial p_i}|(p_i = f_i), \forall i$$

Since $S$ is related with the whole set p instead of only with a specific $pi = f_i$, and the right hand side of the above equations is only determined by the specific $pi = f_i$. This can be true only if $\frac{\partial g^*}{\partial p_i} = 0, \forall i$ and $g^* = 1$. I have completed the proof of theorem 5.

It is obvious that the theorem (5) contains both theorem 1 and theorem 2 and is a very general theorem. With the theorems of 1-5, the mathematical foundation of the constraint-based statistics has been laid. The basic idea of the constraint-based statistics is to determine everything in canonical ensemble including the forms of the entropy and equilibrium distributions with the realistic constraints in nature and with the maximal entropy principle or other types of extremum principles. The application of the constraint-based statistics for non-extensive statistics will be discussed elsewhere.

## VIII. A JUSTIFICATION TO THE FUNDAMENTAL POSTULATE OF STATISTICAL MECHANICS

Given a testable information, the standard maximum entropy procedure developed by Jaynes [14] consists of seeking the probability distribution which maximizes information entropy, subject to the constraints of the information. Information entropy maximization with no testable information takes place under a single constraint: the sum of the probabilities must be one. Under this constraint, the maximum entropy probability distribution is the uniform distribution,

$$f_i = \frac{1}{n}, i = 1,2,...,n$$

The Maximum Entropy Principle (MEP) can thus be seen as a generalization of the classical principle of indifference, also known as the principle of insufficient reason. This generalization is also a justification to the *equal a priori* probability postulate which is fundamental for statistical mechanism. In the following theorem 6, the information entropy is generalized to be target functions satisfying certain weak conditions.

*Theorem 6*

Assume $A$ is a finite set of $\{A_1, A_2..., A_n\}$ and is a source alphabet, $p$ is the probability set of $\{p_1, p_2..., p_n\}$ assigned to $A$, and the probability set $p$ has an achievable state of the set $f$ which can be expressed as $\{f_1, f_2,..., f_n\}$. Then, the equilibrium distribution $f$ should be and should only be uniform distribution or $f_i = \frac{1}{n}, \forall i$, if only natural constraint is put on A or $\sum_{i=1}^{n} p_i = 1$, and if the entropy $S$ as the target function for the extremization with Maximal Entropy Principle (MEP) satisfies two relatively weak conditions:

1) If $\frac{\partial S}{\partial pi}|(p_i = f_i) = constant, \forall i$, then $f_i = \frac{1}{n}$,

2) If $\frac{\partial S}{\partial pi}|(p_i = f_i) = constant, \forall i$, then $S$ becomes maximal.

*Proof of Theorem 6*

The Lagrangian L can be expressed as

$$L = S - \mu \sum_{i=1}^{n} p_i \quad (20)$$

where $S$ is the undetermined entropy, and $\mu$ is underdetermined multiplier. Extremization of Eq. (20) gives the maximal entropy $S$. Setting $\frac{\partial L}{\partial p_i}|(p_i = f_i) = 0, \forall i$ yields the equations resulting in the maximal entropy (equilibrium) distribution $f$:

$$\frac{\partial S}{\partial p_i}|(p_i = f_i) = \mu, \forall i \quad (21)$$

With the above-mentioned weak conditions of 1) and 2), one concludes that $S$ becomes maximal and that the equilibrium distribution is the uniform distribution or $f_i = \frac{1}{n}, \forall i$.

Obviously Boltzmann-Gibbs Entropy satisfies the conditions of 1) and 2), and thus will result in a uniform distribution with the natural constraint. Let us consider another function of $S$

$$S = (n-1) - \sum_{i=1}^{n-1}\sum_{j=i+1}^{n}(p_i - p_j)^2 \qquad (22)$$

The physical meaning of the $S$ in Eq. (22) is that it is a quantity describing the degree of the isotropy of the probability distribution of $\{p_1, p_2 ..., p_n\}$. Since a uniform distribution with full isotropy can be expressed as a super spherical surface in n-dimensional spherical coordinate systems, I called the $S$ in Eq. (22) the "roundness" of the probability distribution of $\{p_1, p_2 ..., p_n\}$. The "roundness" $S$ can be simplified as $S = n(1 - \sum_{i=1}^{n} p_i^2)$.

The simple derivation is as follows

$$S = (n-1) - (n-1)\sum_{i=1}^{n} p_i^2 + 2\sum_{i \neq j} p_i p_j$$

$$= (n-1) - (n-1)((\sum_i p_i)^2 - 2\sum_{i \neq j} p_i p_j) + 2\sum_{i \neq j} p_i p_j$$

$$= n(2\sum_{i \neq j} p_i p_j) = n(1 - \sum_{i=1}^{n} p_i^2)$$

The average "roundness" $S_a$ can thus be expressed [10]

$$S_a = \frac{S}{n} = 1 - \sum_{i=1}^{n} p_i^2$$

It has become clear by now that the average roundness $S_a$ with a clear physical meaning is nothing else but a special Tsallis entropy with q=2. From Eq. (22), one knows that the average roundness $S_a$ as a special Tsallis entropy satisfies conditions 1) and 2), and thus will also determine a uniform distribution with the natural constraint.

It was pointed out that the Boltzmann-Gibbs entropy is also with a similar characteristic like the average "roundness" [10], and can be used to describe the degree of the isotropy.

Some colleagues may think that conclusions based on the equal a priori probability postulate are more fundamental and are with a deeper insights than the results obtained from other approaches including those with the MEP. I would like to argue that the equal a priori probability postulate by itself is nothing else but the result caused by the natural constraint working together with some extremum principles like the MEP.

## IX.  CONCULSION

The ideas of the constraint-based statistics are introduced. The axiom and six theorems of the constraint-based statistics are established. The current main viewpoints of the developing constraint-based statistics are summarized as shown below.

(1) The constraints working together with the extremum principles like maximal entropy principles (MEP) not only fully determine the forms of entropy as the target function but also physically induce the equilibrium distributions.

(2) The constraints contain the essential information about the equilibrium distributions and are a constrained and averaged general information which can be called "pilot information" guiding the evolution towards the equilibrium distributions.

(3) The constraints determining linear functions of Boltzmann-Gibbs entropy as the target functions are a constrained and averaged Weber-type general information, and the constraints forcing the linear functions of Tsallis entropy to be the target functions are a constrained and averaged Stevens-type general information.

(4) The Tsallis entropy with $q = 2$ is with a clear physical meaning and describes the degree of the isotropy of probability distributions.

(5) For classic extensive systems, the Boltzmann-Gibbs entropy or Shannon entropy is directly related with the logarithm of the probability of occurrence [15], the unified constraints expressed by the equation (1) are thus with a maximal probability of occurrence.

## ACKNOWLEDGMENT

I would like to thank Zhang Xuewen and Ouyang Yushan for their understanding and supporting. I also like to thank Dr. R. K. Niven for his interest in this paper.

## REFERENCES

[1]  C. Tsallis, J. Stat. Phys. 52, (1988) 479.
[2]  D.H.E. Gross, arXiv: cond-mat/0210448,v1 21 Oct 2002.
[3]  R. Hanel and S. Thurner, arXiv: cond-mat/0412016 v1 1 Dec 2004.
[4]  S. Abe and A. K. Rajagopal, arXiv:cond-mat/0002159 10 Feb 2000.
[5]  A. Rostovtsev, arXiv:cond-mat/0507414 v1 18 Jul 2005.
[6]  R. Hanel and S. Thurner, arXiv: cond-mat/0602389 v3 24 Jul 2006.
[7]  R. K Niven, arXiv: cond-mat/0503263 v1 11 Mar 2005.
[8]  X.J. Feng, WCFSGS, Vol. 2, No.3, Mar. 2006 (in Chinese)
[9]  X.J. Feng, WCFSGS, Vol. 3, No.4, April 2007 (in Chinese)
[10] X.J. Feng, WCFSGS, Vol. 3, No.2, Feb. 2007 (in Chinese)
[11] GA Gescheider, *Psychophysics: The Fundamentals,* Lawrence Erlbaum Associates,1997.
[12] STEVENS SS, Psychol Rev. 1957 May;64(3):153-81.
[13] G. L. Ferri, S. Martinez, and A. Plastino, J.. Stat. Mech. (2005) P04009, April 20, 2005.
[14] ET Jaynes, Phys. Rev. 106, 620 (1957).
[15] Zhang Xuewen, *The Constitution Theory* (in Chinese), Dec. 2003.